\documentclass[aps,a4paper,12pt,twoside]{revtex4}

\usepackage{float}
\usepackage{fancyhdr}
\usepackage{color}
\usepackage{graphicx}
\usepackage{epsfig} 
\usepackage{amssymb}
\usepackage{amsmath}
\usepackage{booktabs}
\usepackage{hyperref}
\usepackage[footnotesize]{caption}

\def\bra#1{\mathinner{\langle{#1}|}}
\def\ket#1{\mathinner{|{#1}\rangle}}

\def\Bra#1{\left<1>}
{\catcode`\|=\active\gdef\Braket#1{\left<\mathcode`\|"8000\let|\bravert {#1}\right>}}
\def\bravert{\egroup\,\vrule\,\bgroup}

\def\Tr{\mathop{\mbox{\normalfont Tr}}\nolimits}

\begin{document}

\title{Excited state entanglement in homogeneous fermionic chains.}

\author{F. Ares}
\affiliation{Departamento de F\'{\i}sica Te\'orica, Universidad de Zaragoza,
50009 Zaragoza, Spain}
\author{J. G. Esteve}
 \email{esteve@unizar.es}
 \affiliation{Departamento de F\'{\i}sica Te\'orica, Universidad de Zaragoza,
50009 Zaragoza, Spain}
\affiliation{Instituto de Biocomputaci\'on y F\'{\i}sica de Sistemas
Complejos (BIFI), 50009 Zaragoza, Spain}
  \author{F. Falceto\footnote{Corresponding author.} }
\email{ falceto@unizar.es}
 \affiliation{Departamento de F\'{\i}sica Te\'orica, Universidad de Zaragoza,
50009 Zaragoza, Spain}
\affiliation{Instituto de Biocomputaci\'on y F\'{\i}sica de Sistemas
Complejos (BIFI), 50009 Zaragoza, Spain}
\author{E. S\'anchez-Burillo}
\affiliation{Instituto de Ciencia de Materiales de Arag\'on (ICMA), CSIC-Universidad de Zaragoza,
50009 Zaragoza, Spain}


\begin{abstract} 
We study the  R\'enyi entanglement entropy of an interval
in a periodic fermionic chain
for a general eigenstate of a free, translational invariant Hamiltonian. 
In order to analytically compute 
the entropy we use two technical
tools. The first one is used to reduce logarithmically the complexity of 
the problem and the second one to compute the R\'enyi entropy of the
chosen subsystem.
We introduce new strategies to perform the computations, derive new 
expressions for the entropy of these general states and show the perfect 
agreement of the analytical computations and 
the numerical outcome.
Finally we discuss the physical interpretation of our results and
generalise them to compute the entanglement entropy for a fragment of 
a fermionic ladder.
\end{abstract}

\maketitle

\section{Introduction}

In the last two decades, a great effort has been made to quantify the degree of entanglement between two quantum systems. Indeed, there are different ways to measure quantum correlations
\cite{Vedral, Plenio}. One appropriate magnitude which can be used for this purpose is the von Neumann entropy of the reduced state. 
Suppose we have a bipartite system (i.e. it consists of two subsystems $X$ and $Y$) so the Hilbert space can be expressed as the tensor product
$\mathcal{H}=\mathcal{H}_X\otimes\mathcal{H}_Y$ of the Hilbert space of each subsystem. Let $\rho$ be the density matrix which describes the state of the 
system. The von Neumann entropy of the subsystem $X$ is defined as
\begin{equation}\label{1.1} S_1(X)=-\Tr(\rho_X \log\rho_X)\end{equation}
where $\rho_X$ is the reduced density matrix of $X$, i. e. $\rho_X=\Tr_Y(\rho)$ with $\Tr_Y$ denoting the partial trace to the subsystem $Y$. $S_1(X)$ can be interpreted as a measure of the
information which $X$ and $Y$ share. By definition $S_1(X)\geq 0$, 
and the zero value is attained when the
state is separable into the tensor product. 
Besides, if the total system is in a pure
state $\ket{\psi}$, i. e. $\rho=\ket{\psi}\bra{\psi}$, then
$S_1(Y)=S_1(X)$. A generalisation of the von Neumann entropy is the so-called R\'enyi entropy defined for any $\alpha>1$ as
\begin{equation}
\label{1.2}S_\alpha(X)=\frac{1}{1-\alpha}\log\Tr(\rho^\alpha_X).
\end{equation}
In fact when we take the limit $\alpha\rightarrow 1$ we recover the expression (\ref{1.1}). 
Both entropies share the same general properties so that the R\'enyi entropy of the reduced density matrix can also be employed to
measure the entanglement. 

The von Neumann entropy has been specially studied in extended quantum systems
because it is a very suitable quantity to analyse their universal properties 
in the neighbourhood of quantum critical points \cite{Doyon}. In this respect, 
a well-known result is the von Neumann entropy when $X$ is a single interval of length $L$ in a fermionic,
unidimensional chain of $N$ sites, in the ground state of a conformal critical Hamiltonian, with periodic boundary conditions. 
In this case one has \cite{Calabrese, Holzhey, Vidal}
$$S_1(X)=\frac{c}{3} \log\left(\frac{N}{\pi}\sin \frac{\pi L}{N}\right)+C_1\simeq \frac{c}{3}\log L+C_1, \quad N\rightarrow \infty,$$
or for the R\'enyi entropy
\begin{equation}\label{1.3}
S_\alpha(X)=\frac{1+\alpha}{\alpha}\frac{c}{6}\log\left(\frac{N}{\pi}\sin \frac{\pi L}{N}\right)+C_\alpha\simeq 
\frac{1+\alpha}{\alpha}\frac{c}{6}\log L+C_\alpha, \quad
N\rightarrow \infty.
\end{equation}
Here $c$ is the central charge of the underlying conformal field theory
and $C_\alpha$ a non universal constant that will be computed later. 
However, less attention has been paid to the case when the system is in an excited 
state which can strongly change the
behaviour of the entropy \cite{Alba, Sarandy, Alcaraz, Taddia, Dalmonte, Masanes}. In this paper we study the R\'enyi 
entropy for certain particular states, not only the ground state, in a unidimensional
fermionic chain.

In order to carry out the computation we shall make use
of two technical tools. The first one is based on the work of Peschel \cite{Peschel} and 
allows to reduce the dependence of the complexity of the problem 
on the size of the system from $2^L$ to $L$. Instead of studying the 
reduced density matrix it is enough to consider the two-point
correlation matrix.
The second tool can be applied when the correlation matrix
has the Toeplitz form, as it is in our case. 
Then we can find the behaviour of the R\'enyi entropy 
in the thermodynamic limit using the Fisher-Hartwig conjecture 
(or rather theorem, as it has been proven in our case \cite{Basor}). 

The paper extends the works of Jin and Korepin \cite{Jin} and 
Alba, Fagotti and Calabrese \cite{Alba} which employ the previously mentioned
techniques to compute the entanglement in excited states
(see also refs. \cite{Alcaraz, Kadar, Eisler, Keating} for other approaches to the subject). We also generalise these techniques to derive the 
R\'enyi entropy for a piece of a fermionic ladder.

The paper is organised as follows. In the next section
we formulate precisely the problem and fix the notation.
In section \ref{sec3} we discuss the connection between
the reduced density matrix and the two-point correlation matrix
for cases in which the Wick decomposition applies.
In section \ref{sec4} we study the conditions under which
the correlation matrix is of the Toeplitz type and 
the Fisher-Hartwig conjecture holds. Section \ref{sec5} is devoted to the 
analytic evaluation of the R\'enyi entropy for our general case,
while some particular 
examples and its comparison with the numerical results are discussed in section 
\ref{sec6}.
In section \ref{sec7} we address the physical interpretation of our results
and their connection with fermionic chains and ladders. We also compute the 
entanglement entropy for a fragment of a ladder.
Finally, in section \ref{sec8} we present a few conclusions and comments.

\section{Statement of the problem}\label{sec2}

Our system consists in  a chain of $N$ identical, spinless fermions 
with $a_n$ and $a_n^\dagger$,
$n=1,\dots,N$ 
representing respectively the 
annihilation 
and
creation  
operator for the site $n$. 
The only non vanishing anticommutation relations are
$$\{a_n,a_m^\dagger\}=\delta_{nm}.$$

We shall consider states that can be written as a Slater determinant, i.e.
\begin{equation}\label{state}
\vert \Psi_{\cal K} \rangle=  \prod_{k\in {\cal K}} b_k^\dagger \vert 0\rangle
\end{equation}
where $\vert 0\rangle$ represents the vacuum
in the Fock space, 
$$b_k^\dagger=\sum_{n=1}^N \varphi_{k}^*(n)a_n^\dagger,\quad k=1-N/2,\dots,N/2,$$ 
is a basis for the creation operators such that $b_k$ and $b_k^\dagger$ satisfy
canonical anticommutation relations and 
${\cal K}\subset\{1-N/2,\dots,N/2\}$ is the subset of excited modes in $\vert \Psi_{\cal K} \rangle$.

In actual applications, we shall consider a free, translational invariant
Hamiltonian, like that of the Tight Binding Model with periodic boundary conditions
\begin{equation}\label{t-b}
 H= -T\sum_{n=1}^N a^\dagger_n(a_{n-1}+a_{n+1}),
\end{equation}
and the state 
$\vert \Psi_{\cal K} \rangle$
is going to be an eigenstate of the Hamiltonian.
For the moment, however, we will keep the discussion completely
general without making any particular assumptions on the operators $b_k$.

Now we decompose the chain into two subsets $X=\{1,\dots,L\}$ and 
$Y=\{L+1,\dots,N\}$. Adapted to this decomposition we can factor out the Hilbert 
space ${\cal H}={\cal H}_X\otimes{\cal H}_Y$. The goal is to
study the entanglement between the two subsystems. 

In order to do that we introduce the reduced 
density matrix $\rho_X=\Tr_Y(|\Psi_{\cal K}\rangle\langle \Psi_{\cal K}|)$, that
in general does not correspond to a pure state, and
compute its R\'enyi entropy. As it was discussed before
the entropy of the subsystem $X$ coincides with that of the subsystem
$Y$ and provides a measurement for the entanglement between both subsystems.

Once we have obtained the reduced density matrix, we need to
compute its eigenvalues in order to evaluate its R\'enyi entropy.
Considering that the dimension of ${\cal H}_X$ is $2^L$ the 
computational time grows, in principle,  exponentially with the size 
of the subsystem.

As we will see in the next section, there is an algorithm due to
Peschel \cite{Peschel}, that allows to reduce the exponential 
growth to a potential one. 
 
\section{Wick decomposition}\label{sec3}

Wick theorem is a key ingredient in the perturbative expansion of
quantum field theory. It implies that the correlation function of
$2J$ points can be decomposed into the correlation functions
of the different possible pairings of the points. 

We say that a state with density matrix $\rho$ satisfies the 
Wick decomposition 
property if the correlation of an odd number of points is zero
and for every $J$ we have
$$\Tr(\rho d_1\dots d_{2J})=
\frac1{J!}
\sum_{\sigma\in S'_{2J}} \prod_{j=1}^J (-1)^{|\sigma|} \Tr(\rho d_{\sigma(2j-1)} d_{\sigma(2j)}),
$$
where 
$$d_j=\sum_{n=1}^N \alpha_j(n)a_n+\beta_j(n)a^\dagger_n$$
is any linear combination of creation and annihilation operators,
$$S'_{2J}=\{\sigma\in S_{2J}\, |\, \sigma(2j-1)<\sigma(2j),\, j=1,\dots,J\}$$
is the set of permutations that preserve the order in every pair
and $|\sigma|$ is the signature of $\sigma$.
It is interesting to notice that if the original density matrix
of our system $\rho$ satisfies such a property, then the reduced density 
matrix $\rho_X$ also admits the Wick decomposition, 
as one can trivially check.

In the following we will further assume that  $\rho$ preserves the 
total fermionic number and therefore  
$\Tr(\rho a_n a_m)=\Tr(\rho a^\dagger_n a^\dagger_m)=0$. This 
property is also 
inherited by the reduced density matrix.

In order to proceed we must solve the inverse problem. 
That is, given the correlation matrix
$$C_{nm}= \Tr(\rho a^\dagger_n a_m)$$
and assuming that the state satisfies the Wick decomposition 
property, determine the density matrix.

To achieve this goal we diagonalise the correlation matrix
that has real eigenvalues in the interval $[0,1]$. Denote by
$\{\phi_1,\dots,\phi_N\}$ the orthonormal basis of eigenvectors 
with respective eigenvalues $\mu_l$, $l=1,\dots,N$,
and by 
$$c_l=\sum_{n=1}^N\phi_l(n) a_n$$ 
the basis of annihilation
operators for which the correlation matrix is diagonal.
They satisfy, of course, the canonical anticommutation relations.   
We shall consider three disjoint sets in $\{1,\dots,N\}$:
$E_1$ which contains the indices $l$ for which $\mu_l=0$,
$E_2$ that contains the indices such that $\mu_l=1$
and finally $E_3$ that contains the rest of indices, i.e. those
with $0<\mu_l<1$. 

Consider first the set $E_1$. It is clear that if 
$\Tr(\rho c^\dagger_p c_p)=0$, i.e.  $p\in E_1$ and we denote by 
$P_p$ the orthogonal projector into the image of $c_p$, then
we can write $\rho=Q P_p$. By the same token if 
$\mu_q=1$ then $\rho=Q' (1-P_q)$.
Following these remarks, 
we introduce the ansatz
for the density matrix
$$\rho=K{\rm e}^{-h}
\prod_{p\in E_1} P_p
\prod_{q\in E_2} (1-P_q),
$$
with $h=\sum_{l\in E_3}\epsilon^{}_l c^\dagger_l c_l$ and $K$ 
the normalisation constant.
It is easy to see that $\rho$ satisfies the Wick decomposition property, 
the eigenspaces with eigenvalues 0 and 1 in the correlation matrix match and, as we will show below, 
with the adequate choice of $h$ we can also account for the other eigenvalues.

If $\rho$ describes completely the state of the system 
$\Tr(\rho)=1$. Using this fact, we can determine the 
normalisation 
constant
$$K=\frac{1}{\prod_{l\in E_3} (1+e^{-\epsilon^{}_l})}$$
Therefore, the normalised $\rho$ is
\begin{equation}\label{2.0.1}\rho=
\prod_{p\in E_1}P_p \prod_{q\in E_2}(1-P_q)
\prod_{l\in E_3} \frac{e^{-\epsilon^{}_l c_l^\dagger c_l}}{1+e^{-\epsilon^{}_l}}.
\end{equation}
In order to determine the coefficients $\epsilon^{}_l$ we compute
$$\Tr(\rho c^\dagger_l c_l)=\frac1{1+e^{\epsilon^{}_l}},$$
and, then, we can relate the eigenvalues of the correlation matrix to
those of $h$,
\begin{equation}\label{corrham}
\epsilon^{}_l=\log\frac{1-\mu_l}{\mu_l},\quad {\rm for}\quad 0<\mu_l<1.
\end{equation}

If, for a moment, we assume that $E_1$ and $E_2$ are empty, i. e.
$0<\mu_l<1$ for any $l$, then there is a simple relation between $h$ and 
the correlation matrix in terms of the original 
creation and annihilation operator basis. In fact, writing
$$h=\sum_{n,m=1}^N M_{nm}a^\dagger_n a_m$$
and denoting by $C=(C_{nm})$ and $M=(M_{nm})$ the corresponding matrices
one has
$$M=\log(C^{-1}-I)\Leftrightarrow C=({\rm e}^M+I)^{-1}$$

Note that the magic of this procedure is that we have determined
the density matrix, of dimension $2^N$, through a function of
the correlation matrix of dimension $N$.
This fact allows us to go to larger values of $N$ without exhausting
the computational capabilities. Note, however, that this dramatic 
simplification of the problem relies in the fact that the
density matrix is supposed to satisfy the Wick decomposition property,
a fact that must be checked in every case. 
One example that satisfies the Wick decomposition theorem
is the density matrix associated to the Slater determinant
$|\Psi_{\cal K}\rangle$ in (\ref{state}). This case corresponds to $E_3=\emptyset$, 
$E_2={\cal K}$ and $E_1$ its complementary $\cal K^{\rm c}$. Therefore,
if our state of interest $\rho$ satisfies the Wick decomposition property,
according to the previous discussion, also the reduced density matrix
$\rho_X$ has such a property and hence
the previous considerations will be relevant for us.

Once we have determined the density matrix we can compute
its R\'enyi entropy and express it in terms of the eigenvalues
of the correlation matrix. 
In fact, using (\ref{2.0.1}) and (\ref{corrham}) we have
$$\Tr(\rho^\alpha)=\prod_{l\in E_3}\frac{1+{\rm e}^{-\alpha\epsilon^{}_l}}
{(1+{\rm e}^{-\epsilon^{}_l})^\alpha}=\prod_{l=1}^N [(1-\mu_l)^\alpha + \mu_l^\alpha],
$$
where in the last equality the product to {\it all} the eigenvalues of $C$ 
(also those that are 0 or 1) can be 
included without any change in the result.
Therefore
\begin{equation}\label{renyi1}
S_\alpha=\frac1{1-\alpha}\sum_{l=1}^N \log[(1-\mu_l)^\alpha + \mu_l^\alpha],
\end{equation}
expresses the R\'enyi entropy of the density matrix $\rho$
in terms of its correlation. In matrix form we have
$$S_\alpha=\frac1{1-\alpha}\Tr\log[(I-C)^\alpha + C^\alpha].$$

When $\alpha\rightarrow 1$, that is the von Neumann entropy, the previous 
formulae lead to 
$$S_1=-\sum_{l=1}^N \left[(1-\mu_l)\log(1-\mu_l)+\mu_l\log\mu_l\right],$$ 
or 
$$S_1=-\Tr\left[(I-C)\log(I-C)+C\log C\right].$$

The discussion has been restricted to our case of interest in which $\rho$
preserves the total fermionic number and, therefore, 
$\Tr(\rho a_n^\dagger a_m^\dagger)=0$. However a similar result is obtained 
in the general case, with the particularity that the transformation 
to the new creation and annihilation operators $c_l^\dagger, c_l$ that 
diagonalises the correlation matrix is of Bogoliubov type, and the previous 
expressions that involve the correlation matrix should be adequately modified \cite{Peschel}.

\section{Correlation matrix of the Toeplitz type}\label{sec4}

In order to be able to analytically compute 
the entropy of the reduced system
(in the large $N$ limit) it will be crucial that the correlation matrix
is of the 
Toeplitz type, i.e. 
$$C_{nm}=\xi_{n-m}.$$ 
Evidently, this property is a 
consequence of the translational invariance of the system.
We will see now under which conditions it holds. 

Let us consider a state like in (\ref{state})
$$\vert \Psi_{\cal K} \rangle=  \prod_{k\in {\cal K} } b_k^\dagger \vert 0\rangle,$$
where we assume that $b$-operators 
$$b_k=\sum_{n=1}^N \varphi_{k}(n) a_n,\quad k=1-N/2,\dots,N/2$$
satisfy the canonical 
anticommutation relations.
This fact implies that the matrix built
with these coefficients, $\varphi_k(n)$, is unitary (i.e. they form an orthonormal basis). Then the inverse transformation is
$$a_n=\sum_{k=1-N/2}^{N/2}\varphi^*_k(n) b_k.$$

In the previous section we have shown that the R\'enyi entropy of a subsystem 
of the chain can be obtained from the correlation matrix,
$$C_{nm}=\bra{\psi} a_n^\dagger a_m \ket{\psi}.$$
In particular, for the state $\vert \Psi_{\cal K} \rangle$, 
we find that the elements of the correlation matrix are 
\begin{equation}\label{3.2}
C_{nm}=\sum_{k\in \mathcal{K}} \varphi_k(n) \varphi_k^*(m).
\end{equation}
But, if we demand that $C_{nm}$ is a Toeplitz matrix
the $\varphi_k(n)$ must satisfy
\begin{equation}\label{3.2.1}\varphi_k(n)\varphi_k^*(m)=\xi_k(n-m).\end{equation}
That is, the product $\varphi_k(n)\varphi_k^*(m)$ must be invariant
under translations in $n$. The only
functions which constitute an orthonormal basis and satisfy
(\ref{3.2.1}) are
$$\varphi_k(n)=\frac{1}{\sqrt{N}}e^{{2\pi i  k n}/{N}},$$
i.e. the coefficients of the discrete Fourier transform. 
In this case the correlation matrix reads
\begin{equation}\label{3.2.2} C_{nm}=\frac{1}{N}\sum_{k\in\mathcal{K}} 
e^{{2\pi i k(n-m)}/{N}}.\end{equation}

In the following section it will be convenient to use $V=2C-I$
instead of the correlation matrix. In our case it has the following 
expression
\begin{equation}\label{Vnm} 
V_{nm}=\frac{1}{N}\left(
\sum_{k\in\mathcal{K}} e^{{2\pi i k(n-m)}/{N}}
-
\sum_{k\not\in\mathcal{K}} e^{{2\pi i k(n-m)}/{N}}\right).
\end{equation}

Coming back to the correlation matrix we would like to study some of its transformation properties
that result in the invariance of the entropy.
\begin{enumerate}
 \item Translational invariance of the entropy in the coordinate space ($a$-particles), which comes from the fact that $C$ is a Toepitz matrix. 
 \item Translational invariance in the momentum space ($b$-modes) of the state. Let $\mathcal{K}$ be a particular configuration of occupied modes and let 
$\mathcal{K}'=\mathcal{K}+\Delta=\{k+\Delta | k\in {\cal K}\}$.
The correlation matrix for $\mathcal{K}'$ is 
$$C'_{nm}=\frac{1}{N}\sum_{k'\in\mathcal{K}'} e^{{2\pi i k'(n-m)}/{N}}$$
Changing the variable of the sum, $k=k'-\Delta$
\begin{equation}\label{3.3} C'_{nm}=\frac{1}{N}\sum_{k\in \mathcal{K}'-\Delta} e^{{2\pi i (k+\Delta)(n-m)}/{N}}=e^{{2\pi i n\Delta }/{N}}C_{nm}e^{-{2\pi im\Delta }/{N}}\end{equation}
Defining now the unitary matrix $U$
$$U_{nm}=
e^{-{2\pi i n\Delta }/{N}}\delta_{nm}$$
we can rewrite the expression (\ref{3.3})
$$C'_{nm}=(U^\dagger C U)_{nm}$$
Therefore $C'$ results from applying a unitary transformation to $C$ , so that their eigenvalues are the same and the associated entropy is equal. 
\item Invariance under complementarity in $b$-space. Consider a particular configuration $\mathcal{K}$ and call ${\cal K}^c$ its complemetary set.
The correlation
 matrix for $\mathcal{K}^c$ is
$$C^c_{nm}=\sum_{k\in\mathcal{K}^c}\varphi_{k}(n)\varphi^*_{k}(m)=\frac{1}{N}\sum_{k\in\mathcal{K}^c}e^{{2\pi i k(n-m)}/{N}}$$
Using the fact $\{\varphi_{k}(n)\}$ is an orthonormal basis 
$$\sum_{k=1}^N \varphi_{k}(n)\varphi_k^*(m)=\delta_{nm}$$
we can decompose the last sum into two pieces
$$\delta_{nm}=\sum_{k\in\mathcal{K}}\varphi_{k}(n)\varphi_k^*(m)+\sum_{k\in\mathcal{K}^c}\varphi_{k}(n)\varphi_k^*(m)=C_{nm}+C^c_{nm}.$$
Therefore, the eigenvalues of $C^c_{nm}$ are $\mu^c_l=1-\mu_l$. Inserting them into (\ref{renyi1}) we find that 
the entropies of the states determined by $\mathcal{K}^c$ and $\mathcal{K}$ 
are the same. 
\end{enumerate}

It is interesting to remark that these properties reflect the 
duality between the representation in coordinate space $a_n$ and in 
momentum space $b_k$. In fact in both scenarios we have translational 
invariance of the entropy and also invariance under complementarity.

\section{Entanglement entropy for a single block}\label{sec5}

Suppose that the total correlation matrix of the chain is Toeplitz and a subsystem $X$ which is a single block of $L$ contiguous sites in the coordinate space. In this case, the correlation matrix of the
subsystem, which we denote by $C(X)$, is a Toeplitz matrix as well. 
Using the properties of the Toeplitz matrices, the existence of a
generating function and the Fisher-Hartwig conjecture, we can find how 
the entropy scales with the size of $X$ in the thermodynamic limit. 

If we denote by $v_l=2\mu_l-1$ the eigenvalues of the matrix $V(X)=2\,C(X)-I_L$ (which is a Toeplitz matrix too), the R\'enyi entropy (\ref{renyi1}) can be
written
\begin{equation}\label{3.4}S_\alpha(X)=\frac{1}{1-\alpha}\sum_{l=1}^L\log \left[\left(\frac{1+v_l}{2}\right)^\alpha+\left(\frac{1-v_l}{2}\right)^\alpha\right].\end{equation}
It will be convenient to transform the sum in the previous expression into a complex integral
on a contour that encloses the eigenvalues of $V(X)$. To that end we introduce the function
$$f_\alpha(x,y)=\frac{1}{1-\alpha}\log \left[\left(\frac{x+y}{2}\right)^\alpha+\left(\frac{x-y}{2}\right)^\alpha\right],$$
and applying the Cauchy's residue theorem
we can rewrite the R\'enyi entropy (\ref{3.4}) in terms of a complex integral:
\begin{equation}\label{3.5}S_\alpha(X)=\lim_{\varepsilon\to 0^+}\frac{1}{2\pi i}\oint_{\mathcal{C}}
f_\alpha (1+\varepsilon, \lambda)\frac{d \log D_L(\lambda)}{d\lambda}d\lambda, 
\end{equation}
where $D_L(\lambda)$ is the determinant of $\lambda I_L-V(X)$,
$$D_L(\lambda)=\prod_{l=1}^L (\lambda-v_l),$$
and the contour $\mathcal{C}$ surrounds all the eigenvalues 
of $V$ which are the poles of the integrand.
The contour of integration, the poles and the cuts are depicted in the figure. 

  \begin{figure}[H]
  \centering
    \resizebox{12cm}{4cm}{\includegraphics{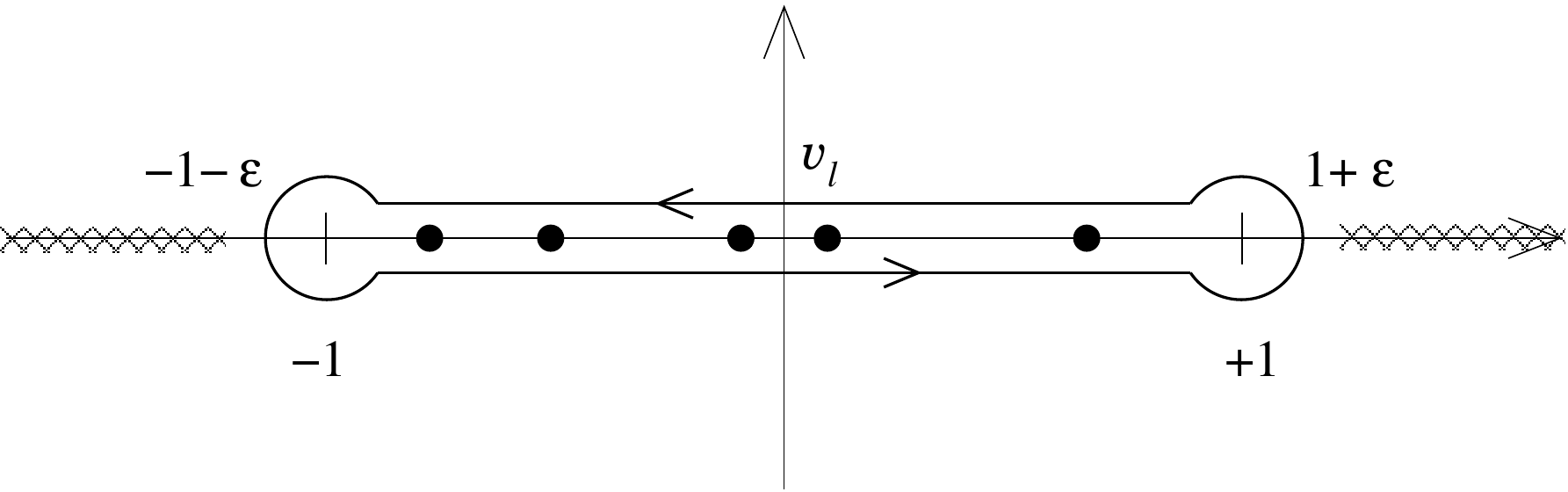}} 
    \caption{Contour of integration, cuts and poles for the computation of 
$S_\alpha(X)$. The cuts for the function $f_\alpha$ extend to $\pm\infty$.}
  \label{contorno0}
   \end{figure}

A state $|\Psi_{\cal K}\rangle$ is characterised by the set of occupied modes 
$\cal K$.
In the thermodynamic limit this can be
approximated by an occupation density, so we describe the state 
through a periodic function  $g(\theta)$ that takes values in the interval 
$[-1,1]$ 
and is defined by
$$\frac{1}{2\pi}\int_{-\pi}^{\pi} F(\theta)g(\theta)d\theta=\lim_{N\to\infty} \frac{1}{N} \left[ 
\sum_{k\in{\cal K}}F\left(\frac{2\pi k}{N}\right)
-
\sum_{k\not\in{\cal K}}F\left(\frac{2\pi k}{N}\right)\right],
$$
where the equality should hold for any continuous function $F$.
Notice that $g(\theta)= 1$ if the modes with momenta around 
$k_\theta=N\theta/(2\pi)$ 
are all occupied and $g(\theta)$ is $-1$ if they are empty. Intermediate values represent the occupation of only a fraction of the $b$-modes 
with momenta near $k_\theta$.

Now, we can replace in (\ref{Vnm}) the sum with an integral 
$$V_{nm}=\frac{1}{2\pi}\int_{-\pi}^{\pi}g(\theta)e^{i(n-m)\theta}
d\theta.$$

The kind of occupation densities we will be interested in are 
piecewise constant functions.  
If we denote by $\theta_1,\dots, \theta_R$ the discontinuity points,
\begin{equation}\label{occupation}
g(\theta)=t_r,\quad \theta_{r-1}\leq\theta<\theta_r.
\end{equation} 
Note that the ground state of a quadratic, translational invariant Hamiltonian with short range interactions, corresponds to $t_r$ taking the values $+1$ or $-1$.
Later we will discuss the physical interpretation of more general occupation 
densities.

Following Jin and Korepin \cite{Jin}, we use the Fisher-Hartwig conjecture to compute the determinant $D_L$ \cite{Fisher, BasorMorrison}.
We express $\tilde{g}(\theta)=\lambda-g(\theta)$ 
in the form
$$\tilde{g}(\theta)=\psi(\theta) \prod_{r=1}^R \tau_r(\theta-\theta_r),$$
where
$$\psi(\theta)=\prod_{r=1}^R (\lambda-t_{r-1})^\frac{\theta_r-\theta_{r-1}}{2\pi}$$
and the discontinuties are taken into account by the functions
$$\tau_r(\theta)= e^{-i\beta_r(\pi-\theta)},\quad \theta\in[0,2\pi),$$
that have a jump at $\theta=0$, while
$$\beta_r=-\frac{1}{2\pi i}\log\left(\frac{\lambda-t_{r-1}}{\lambda-t_r}\right).$$

The Fisher-Hartwig conjecture (which has been proven for our case \cite{Basor}) states that up to a factor that goes to $1$ when $L\rightarrow \infty$, 
the determinant of our Toeplitz matrix, $D_L$, can be approximated by
$$D_L\approx(F[\psi])^L\left(\prod_{r=1}^R L^{-\beta_r^2}\right)E[\{\beta_r\},\{\theta_r\}],$$
where $F[\psi]=e^{\frac{1}{2\pi}\int_{0}^{2\pi} \log (\psi(\theta))d\theta}$ and
\begin{equation}\label{3.6} 
E[\{\beta_r\}, \{\theta_r\}]=\prod_{r=1}^R G(1+\beta_r)G(1-\beta_r)
\prod_{1\leq r\neq r'\leq R}(1-e^{i(\theta_r-\theta_{r'})})^{\beta_r \beta_{r'}}
\end{equation}
with $G(z)$ the Barnes G-function. The logarithmic derivative of $D_L$ is 
$$\frac{d\log D_L(\lambda)}{d\lambda}=\frac{d\log F[\psi]}{d\lambda}L-\sum_{r=1}^R \frac{d\beta_r^2}{d\lambda}\log L +\frac{\log E}{d\lambda}+\dots,$$
where the dots stand for terms that vanish in the large $L$ limit.
Inserting this expression into (\ref{3.5}) we can 
compute the entropy for large $L$. Namely
\begin{equation}\label{expansion}
S_\alpha(X)= A_\alpha L+B_\alpha \log L+C_\alpha+\dots
\end{equation}
and our next task is to compute the coefficients in the expansion.

With respect to the linear term $A_\alpha$, 
if we consider that $\psi$ does not depend on $\theta$, 
we have $F[\psi]=\psi$. Taking its logarithmic derivative, introducing it in (\ref{3.5}) and applying the Cauchy's residue theorem we obtain
\begin{equation}\label{3.7.1}
A_\alpha=\frac{1}{2\pi}\int_{-\pi}^{\pi} f_\alpha(1,g(\theta))d\theta.
\end{equation}

As for the coefficient of the logarithmic term $B_\alpha$, we have 
$$B_\alpha=-\frac{1}{2\pi i}\sum_{r=1}^R \lim_{\varepsilon\to0^+}\oint_\mathcal{C} f_\alpha(1+\varepsilon,\lambda)\frac{d\beta_r^2}{d\lambda}d\lambda,
$$
where  the poles and cuts of the $r$-th term in the sum are 
represented in the figure. 

  \begin{figure}[H]
  \centering
    \resizebox{12cm}{4cm}{\includegraphics{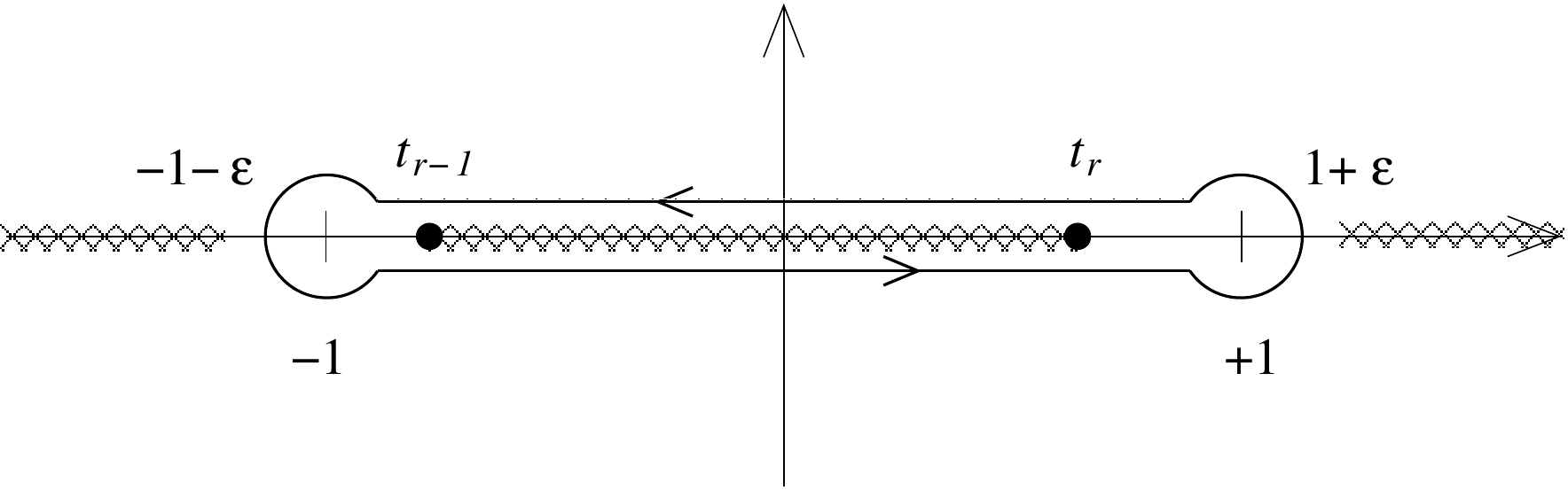}} 
    \caption{Contour of integration, cuts and poles for the computation of 
$B_\alpha$.}
  \label{contorno}
   \end{figure}

This integral has been considered before in the literature. 
The usual strategy to 
compute it goes through the decomposition of the contour into different pieces
with divergent integrals.
After an annoying computation it is shown that the divergences 
cancel to render a finite result. See \cite{Jin} for the original work
and \cite{Eisler} for a recent application.
We propose a new strategy that leads directly to the finite result 
in a simpler way. It consists in performing an integration by parts, 
so that
$$
B_\alpha=\frac{1}{2\pi i}\sum_{r=1}^R
\lim_{\varepsilon\to0^+}
\oint_\mathcal{C}
\frac{df_\alpha(1+\varepsilon,\lambda)}{d\lambda} \beta_r^2d\lambda.
$$
Now a straightforward computation of the integral along $\cal C$ gives
$$B_\alpha= \frac{1}{2\pi i}\sum_{r=1}^R\frac{1}{4\pi^2}\int_{t_{r-1}}^{t_r} \frac{df_\alpha(1,\lambda)}{d\lambda}
\left[\left(\log\left|
\frac{\lambda-t_r}{\lambda-t_{r-1}}
\right|+i\pi\right)^2-\left(\log\left|
\frac{\lambda-t_r}{\lambda-t_{r-1}}
\right|-i\pi\right)^2\right]d\lambda,$$
and at last
\begin{equation}\label{Ba}
 B_\alpha=\frac{1}{2\pi^2}\sum_{r=1}^R\int_{t_{r-1}}^{t_r}  
\frac{df_\alpha(1,\lambda)}{d\lambda}
\log\left|\frac{\lambda-t_r}{\lambda-t_{r-1}}\right|
d\lambda.
\end{equation}
So, we write the logarithmic coefficient as an integral that, in general, 
should be determined numerically. In the next section we will give some
examples in which $B_\alpha$ is analytically computed. 

Finally, we must obtain the finite contribution $C_\alpha$. 
Proceeding as before we have
\begin{multline*}
C_\alpha=\frac{1}{2\pi i}
\lim_{\varepsilon\to 0^+}\oint_{\mathcal{C}} f_\alpha(1+\varepsilon,\lambda)
\left[\sum_{r=1}^R\frac{d}{d\lambda}\log[G(1+\beta_r)G(1-\beta_r)]\right.+ 
\\ +\left.\sum_{1\leq r\neq r'\leq R}
\log(1-e^{i(\theta_r-\theta_{r'})})\frac{d}{d\lambda}(\beta_r\beta_{r'})\right]
d\lambda 
\end{multline*}

We have to deal with two different integrals. 
The first one reads
\begin{eqnarray}
I_\alpha(r)&=&\frac{1}{2\pi i}\lim_{\varepsilon\to 0^+}\oint_{\mathcal{C}} 
f_\alpha(1+\varepsilon, \lambda)
\frac{d}{d\lambda}
\log[G(1+\beta_r)G(1-\beta_r)] 
d\lambda \cr
&=&
\frac{-1}{2\pi i}
\lim_{\varepsilon\to 0^+}
\oint_{\mathcal{C}} 
\frac{df_\alpha(1+\varepsilon,\lambda)}{d\lambda}
\log[G(1+\beta_r)G(1-\beta_r)]
d\lambda 
\end{eqnarray}
where we have performed an integration by parts.
The cuts in the complex plain for the above integral are again those of 
fig. \ref{contorno}. Therefore, after integrating along the contour we get
$$I_\alpha(r)=\frac{1}{2\pi i}\int_{t_{r-1}}^{t_r} \frac{df_\alpha (1,\lambda)}{d\lambda}
\log\left[\frac{G(1+\beta_r^-)G(1-\beta_r^-)}{G(1+\beta_r^+)G(1-\beta_r^+)}\right]d\lambda$$
where
$$
\beta_r^\pm =i\omega_r(\lambda)\pm\frac{1}{2},
\quad {\rm with}\quad
\omega_r(\lambda)=\frac{1}{2\pi}\log\left|\frac
{\lambda-t_r}{\lambda-t_{r-1}}
\right|.
$$
Note that $\beta_r^+=\beta_r^-+1$, so applying the property of the 
Barnes G-function $G(z+1)=\Gamma(z) G(z)$, where $\Gamma$ is the 
Euler gamma function, we finally obtain
\begin{equation}\label{I1}
 I_\alpha(r)=\frac{1}{2\pi i}\int_{t_{r-1}}^{t_r} \frac{d f_\alpha(1,\lambda)}{d\lambda}
 \log\left[\frac{\Gamma(1/2-i\omega_r(\lambda))}{\Gamma(1/2+i\omega_r(\lambda)}\right]d\lambda.
\end{equation}

The computation of the second integral
goes along similar lines. Define
\begin{eqnarray*}
  J_\alpha(r,r')&=&-\frac1{2\pi i}
\lim_{\varepsilon\to 0^+}\oint_{\mathcal{C}} f_\alpha(1+\varepsilon, \lambda)\frac{d(\beta_r \beta_{r'})}{d\lambda}d\lambda \cr
&=&\frac{1}{2\pi i}\lim_{\varepsilon\to 0^+}\oint_{\mathcal{C}}
\frac{df_\alpha(1+\varepsilon,\lambda)}{d\lambda} \beta_r \beta_{r'} d\lambda, 
\end{eqnarray*}
where, again, we find useful to perform an integration by parts
to achieve the cancellation of divergences. 
There are two cuts in the region inside the integral contour going from 
$t_{r-1}$ to $t_{r}$ 
and from $t_{r'-1}$ to $t_{r'}$
respectively. The integral should be decomposed into different regions
corresponding to the different cuts and computed carefully.
But, considering that we only meet harmless logarithmic singularities,
we can appropriately group the different terms to obtain
$$J_\alpha(r,r')=K_\alpha(r,r')+K_\alpha(r',r)$$
where
$$K_\alpha(r,r')=\frac{1}{4\pi^2}\int_{t_{r-1}}^{t_r} \frac{df_\alpha(1,\lambda)}{d\lambda}
\log\left|
\frac{\lambda-t_{r'}}{\lambda-t_{r'-1}}
\right| d\lambda$$

Therefore, putting all together we have 
\begin{equation}
C_\alpha=\sum_{r=1}^{R} I_\alpha(r)
-
\sum_{1\leq r\neq r'\leq R}\log[2-2\cos(\theta_r-\theta_{r'})] K_\alpha(r,r')
\end{equation}
To our knowledge, this is the first time that such an explicit expression of
the constant term, for general values of $t_r$, appears in the literature.

Note that if our state has $g(\theta)=\pm 1$, with a finite number of discontinuities, the entropy of $X$ scales with $\log L$ because in this case the coefficient of the linear term vanishes. 
It has been shown in  \cite{Alba} that these configurations correspond to the 
ground state of a Hamiltonian with local interactions. 
On the other hand, if there are intervals with $g(\theta)\neq \pm 1$ 
the entropy exhibits both linear and logarithmic contributions. 
It has been proposed that these configurations correspond, in general,
to the ground state of a non-local Hamiltonian \cite{Alba}.
In section \ref{sec7} we will carry out a thorough discussion on
the connection of our kind of states, its entanglement entropy and 
quadratic Hamiltonians for chains and ladders.

\section{Numerical results}\label{sec6}
In this section we present the results  
for the R\'enyi entanglement entropy of a single 
interval of length $L$, when $\alpha\rightarrow 1$ (von Neumann entropy) and
$\alpha=2$.
We perform the computation using the analytic expressions obtained in the previous section for the expansion in $L$ and compare with the numerical results. 
As we discussed before, thanks to the Wick decomposition property 
\cite{Peschel}, we can easily cover values of $L=10^3$, in the numeric computation (note  that $\rho_X$ acts on Hilbert space of dimension $2^{1000}$ which is impossible to deal with in any computer).

We consider three different pure states, each one characterised by a set 
$\cal K$ of occupied $b$-modes. In the thermodynamic limit
they are described by its occupation density. 
With these three states we can check different aspects of the 
analytic expressions obtained before.

 \begin{description}
  \item[State 1:] half of the $b$-modes, corresponding to the 
lowest absolute value of the momenta, are 
occupied, while the others are empty. The state can be represented
by $(0\dots01\dots11\dots10\dots0)$ where every digit represents 
the occupation -1- or not -0- of the mode
of the corresponding momentum $k$.
The occupation density is
       $$g(\theta)=\left\{\begin{array}{ll} -1,\;\;\;\;\; & \theta\in\left[-\pi,-{\pi}/{2}\right)\cup\left[{\pi}/{2},\pi\right],\\
                                           1, & \theta\in\left[-{\pi}/{2},{\pi}/{2}\right).
                          \end{array}\right. $$
   Note that this is precisely the ground state of the Tight Binding Model 
with the local periodic Hamiltonian
   \begin{equation}\label{4.1}H=-T\sum_{n=1}^N a^\dagger_n(a_{n-1}+a_{n+1}),\end{equation}
   and $T>0$. Since it is invariant under translations it is diagonalised by the Fourier 
transformation so that its eigenstates are $\ket{\Psi_\mathcal{K}}=\prod_{k\in\mathcal{K}}
b^\dagger_k \ket{0}$
   $$H\ket{\Psi_\mathcal{K}}=\sum_{k\in{\cal K}}\Lambda_k \ket{\Psi_{\mathcal{K}}}; \quad \Lambda_k= -2T\cos\left(\frac{2\pi k}{N}\right)$$
  
  \item[State 2:] $b$-modes with momenta $|k|>N/4$ are empty
while those with $|k|<N/4$ are occupied alternatively.
This can be represented by $(0\dots0\ 10\dots10\ 0\dots0)$. Then the occupation density is
       $$g(\theta)=\left\{\begin{array}{ll} -1,\;\;\;\;\; & \theta\in\left[-\pi,-{\pi}/{2}\right)\cup\left[{\pi}/{2},\pi\right],\\
                                           0, & \theta\in\left[-{\pi}/{2},{\pi}/{2}\right).
                          \end{array}\right. $$
  \item[State 3:] for $|k|>N/4$ all $b$-modes are empty, while for the other half there are three out of four occupied, i. e. 
$(0\dots0\ 1110\dots1110\ 0\dots0)$. Therefore, the occupation density is 
 $$g(\theta)=\left\{\begin{array}{ll} -1,\;\;\;\;\; & \theta\in\left[-\pi,-{\pi}/{2}\right)\cup\left[{\pi}/{2},\pi\right],\\
                                           1/2, & \theta\in\left[-{\pi}/{2},{\pi}/{2}\right).
                          \end{array}\right. $$
  
 \end{description}

\begin{figure}[H]
   \centering
   \includegraphics[width=\textwidth]{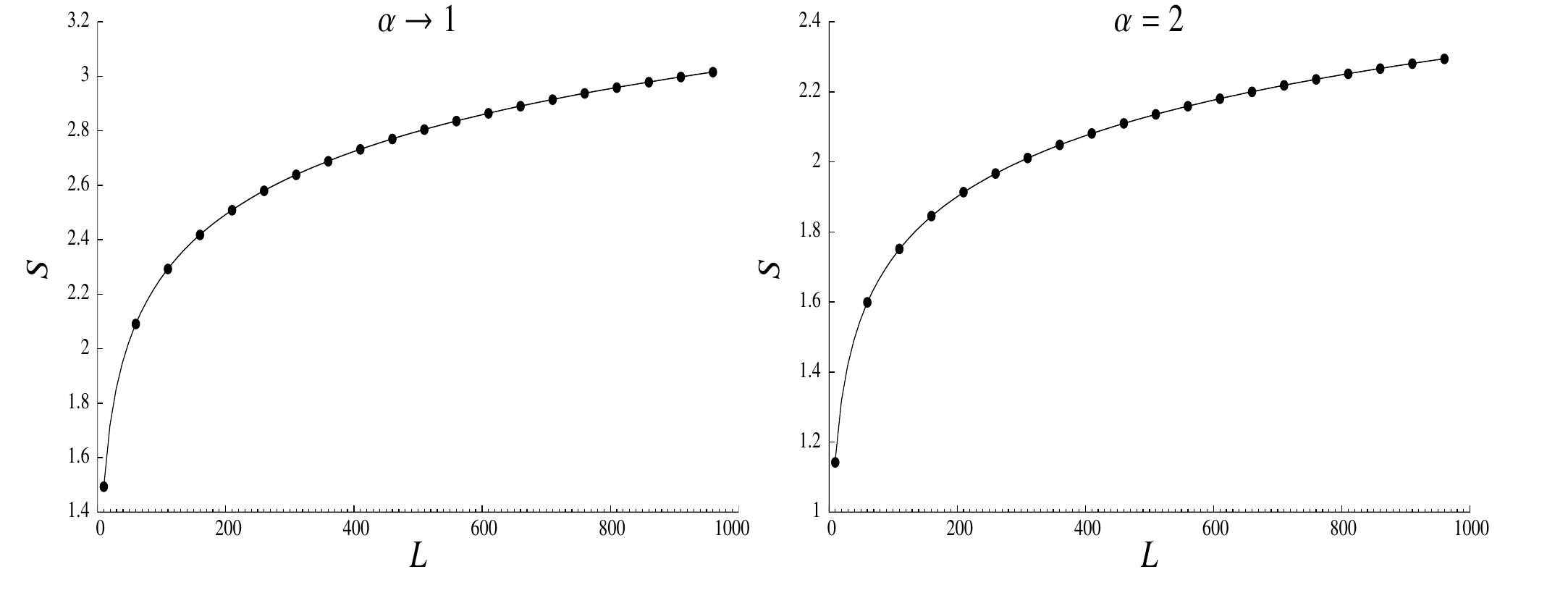}
   \caption{Entropy when $\alpha\rightarrow 1$ (left panel) and $\alpha=2$  
(right panel)  as a function of the length $L$ of the block 
when the chain is in the \textbf{state 1}. The dots are the numerical values 
while the continuous line represents the asymptotic expansion 
(\ref{expansion}) with $A_\alpha=0$, $B_1=1/3$, $C_1=0.726067$ and $B_2=1/4$, $C_2=0.577336$ respectively.}
  \label{state1}\end{figure}

\begin{figure}[H]
  \centering
  \includegraphics[width=\textwidth]{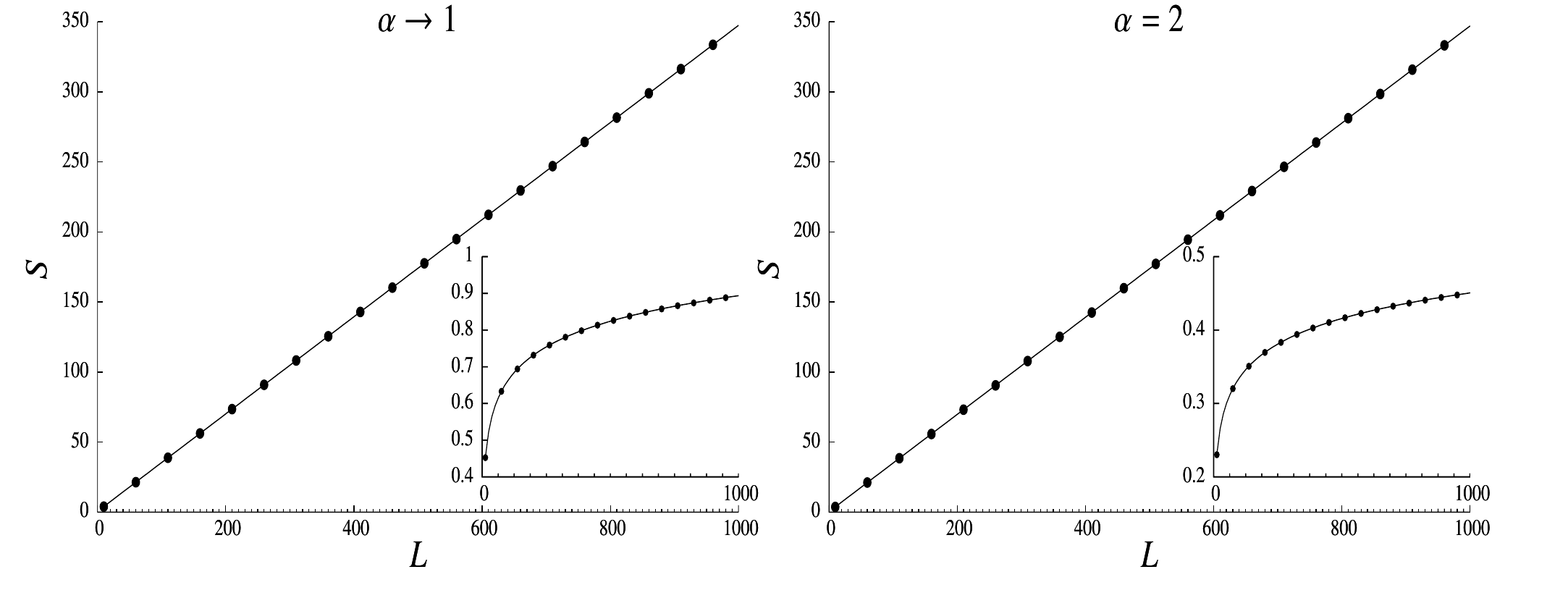}
  \caption{Entropy when $\alpha\rightarrow 1$ (left panel) and $\alpha=2$ (right panel) for the 
\textbf{state 2}. The dots are the numerical values and the continuous line represents the 
Fisher-Hartwig expansion (\ref{expansion}) with 
the coefficients of table 1.
In the insets we plot the entropy subtracting the
linear contribution to reveal the logarithmic term.}
  \label{state2}\end{figure}

\begin{figure}[H]
  \centering
  \includegraphics[width=\textwidth]{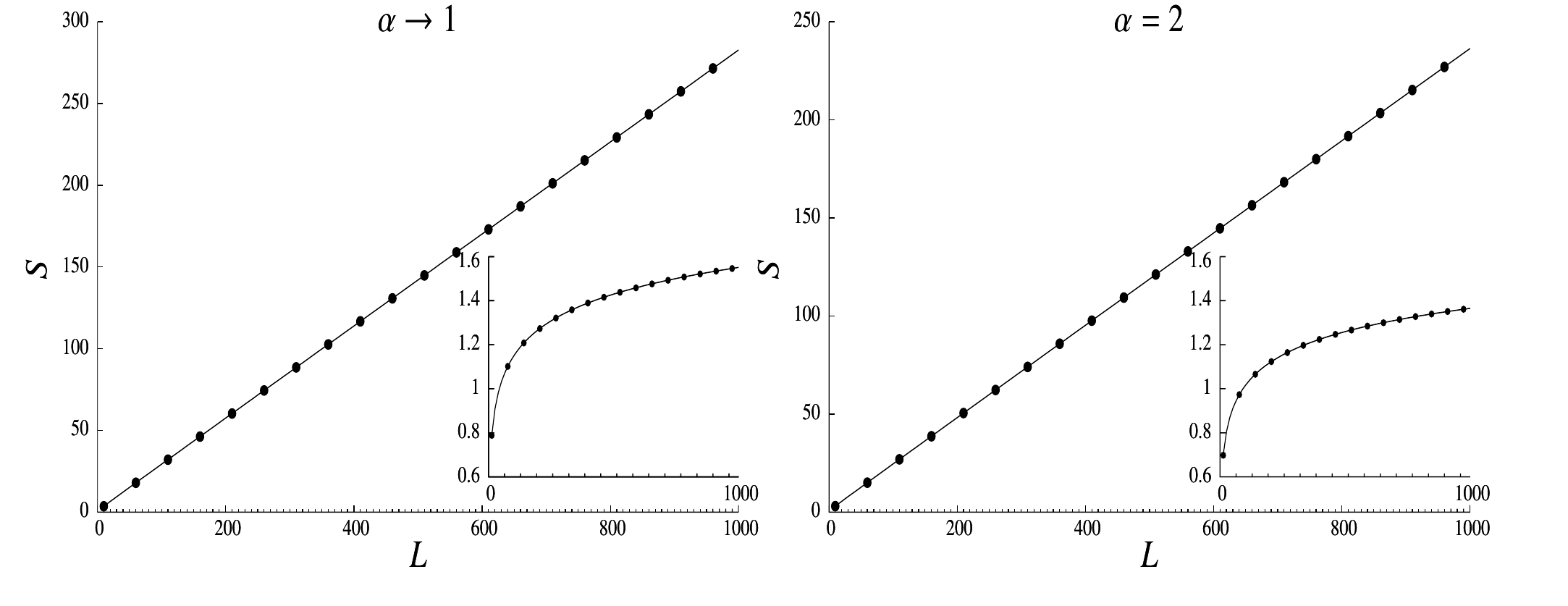}
  \caption{Entropy when $\alpha\rightarrow 1$ (left panel) and $\alpha=2$ (right panel) in the \textbf{state 3}. 
The dots are the numerical values while the continuous line 
represents the Fisher-Hartwig expansion (\ref{expansion}) 
with the coefficients of table 1.
In the insets we plot the entropy subtracting the linear
contribution.}
  \label{state3}\end{figure}

In our numerical calculations we take the thermodynamic limit and 
compute the entropies of a single block with a length varying from $L=10$ to 
$L=1000$ sites. Dots in figures \ref{state1}, \ref{state2} and \ref{state3} 
represent the numerical results, while the continuous line stands for
the analytic results for the large $L$ expansion.  
 
The table \ref{table1} summarises the numerical values of $A_\alpha$, $B_\alpha$ 
and $C_\alpha$ that we get using the expressions found in the 
previous section for these particular states. We
compare them to those obtained by fitting the numerical entropies, 
computed from the correlation matrix of the subsystem.

\begin{table}
 \centering
 \begin{tabular}{|c| c c| c c| c c|}
   \toprule
   {$\alpha\to 1$} & \multicolumn{2}{c|}{$A_1$} & \multicolumn{2}{c|}{$B_1$} & \multicolumn{2}{c|}{$C_1$} \\
   \hline

    State   &     F-H    & num.      &    F-H &    num.     &  F-H      &    num.   \\
    \hline
    \midrule
     1   &       0    & 0         &   1/3     & 0.333343 & 0.726067  & 0.726600 \\

     2   &\,  0.3465735\, &\, 0.346574 \, &\, 0.100660\,  &\, 0.100669  \,  & \,0.220813\,  &\, 0.220768\,  \\

     3   &  0.2811676 & 0.281168  & 0.175015  & 0.175019    & 0.385367  & 0.385460  \\
  \bottomrule
 \end{tabular}

 \vskip 0.5cm
 \begin{tabular}{|c| c c| c c| c c|}
   \toprule
   {$\alpha=2$} &  \multicolumn{2}{c|}{$A_2$} & \multicolumn{2}{c|}{$B_2$} & \multicolumn{2}{c|}{$C_2$} \\
   \hline
 State   &     F-H    & num.      &    F-H    & num.     &  F-H      &    num.   \\
 \hline
   \midrule
     1   &    0       & 0   &    1/4    & 0.251098       & 0.577336  & 0.570234 \\

     2   &\,  0.3465735\, &\, 0.346574\,  &\, 0.050330\,  & \,0.050314 \,  &\, 0.114183 \, & \,0.114270 \, \\

     3   &  0.2350018 & 0.235001  & 0.152477  & 0.152875   & 0.349182  & 0.347217  \\
  \bottomrule
\end{tabular} 

\caption{The ``F-H'' columns collect the predicted values of $A_\alpha$, $B_\alpha$ and $C_\alpha$ using the 
Fisher-Hartwig conjecture when $\alpha\rightarrow 1$ (upper table) and $\alpha=2$ (lower table). 
The corresponding ``num.'' columns contain the value of these coefficients when we fit the numerical 
entropies with the expansion (\ref{expansion}).} 
\label{table1}\end{table}
The agreement between the outcomes is excellent. Note that when the linear term is nonzero, it provides 
the best fit to the numerical entropies because it dominates. On the contrary, the adjustment
of the finite term is the poorest. 
From these results we can conclude that the large $L$ expansion (\ref{expansion})  works rather well 
from lengths of $L=10$ sites.
On the other hand, one may ask how our results are affected
by finite-size effects.
In this sense we have performed numerical simulations
for finite $N$ and we can conclude that the expansion
works reasonably well (when $L\in[10,1000]$) 
from around $N=10^4$.

State 1 is specially interesting, as it corresponds to the ground state of 
(\ref{4.1}) that is a critical (gapless) theory. The entropy grows 
logarithmically as it is predicted from conformal field theory (\ref{1.3}) 
with central charge $c=1$. 
The comparison with the expression for the coefficient of the 
logarithmic term (\ref{Ba}) and (\ref{1.3}) leads to the following 
integral identity,
\begin{equation}\label{intid}\frac{1}{1-\alpha}\int_{-1}^1 
\log\left[
\left(\frac{1+x}2\right)^\alpha
+
\left(\frac{1-x}2\right)^\alpha
\right]
\frac{dx}{1-x^2} = \frac{\pi^2}{12} \frac{1+\alpha}\alpha 
\end{equation}
which can be analytically derived 
with the change of variables 
$x=(t-1)/(t+1)$ that also reveals the relationship with 
the dilogarithm function. 

Concerning the state 2, first note that its linear coefficient does not depend on 
$\alpha$. Actually it is easy to compute to give 
$$A_\alpha=\frac{1}{2\pi}\int_{-\pi/2}^{\pi/2} f_\alpha(1,0) d\theta
=\log(2)/2,$$
that agrees with  the numerical results. In addition, for this configuration
the logarithmic coefficient for integer $\alpha$ 
can be analytically obtained. It has the expression 
$$B_\alpha=\frac{\alpha+1}{24\alpha}-\frac1{2\pi^2(\alpha-1)}
\sum_{n=1}^\alpha\left(\log\sin\frac{(2n-1)\pi}{2\alpha}\right)^2,$$
for $\alpha\geq2$, while for $\alpha=1$ is 
$$B_1=\frac1{8} - \frac12 \left(\frac{\log2}{\pi}\right)^2,$$
values that agree with the numerical ones that  
appear in the table.

\begin{figure}[H]
  \centering
    \resizebox{8cm}{!}{\includegraphics{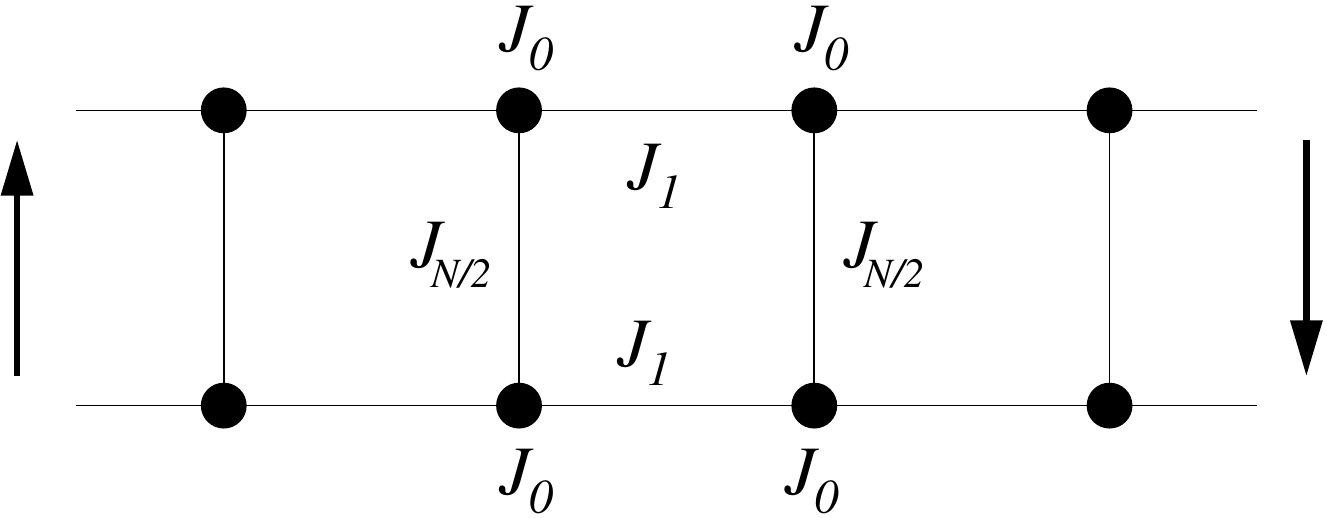}} 
    \caption{Representation of the fermionic ladder with the couplings of 
the Hamiltonian
of (\ref{nonlocalH}). The upper row contains the sites from 1 to $N/2$ 
of the chain and the lower one those form $N/2+1$ to $N$, 
both from left to right.
The ends of the ladder are joined with an inversion as it is 
indicated by the arrows, forming,
therefore, a Moebius strip.
}
  \label{ladder}
   \end{figure}

As it is discussed in \cite{Alba}, states whose entropy grows linearly with the size
of the interval may correspond to the vacuum of a non-local Hamiltonian.
This can be shown to hold in our case. In fact, consider the simple Hamiltonian
\begin{equation}\label{nonlocalH}
H=\sum_n (J_0 a^\dagger_n a_n+J_1 a^\dagger_n a_{n+1}+ J_{N/2} a^\dagger_n a_{n+N/2}) + h. c.,
\end{equation}
that corresponds to a fermionic ladder as it is depicted in the figure \ref{ladder}
(note that one must perform an inversion when joining the two ends
forming a Moebius strip rather than a periodic ladder).
If we take the coupling constants 
$$J_{N/2}=J_0>0,\quad J_1=-2J_0,$$
the ground state is precisely the state 2.

A similar result is obtained for state 3.
If we consider the Hamiltonian 
$$H=\sum_n (J_0 a^\dagger_n a_n+J_1 a^\dagger_n a_{n+1}+ J_{N/4} a^\dagger_n a_{n+N/4} 
+ J_{N/2} a^\dagger_n a_{n+N/2}) + h. c.,$$
with 
$$J_{N/2}=J_0>0,
\quad J_{N/4}=2J_0\quad {\rm and}\quad J_1=-4J_0,$$
its vacuum is state 3. 

Note that our results follow the {\it rule of thumb} according to which
the ground state entanglement entropy grows like the number of bonds that
are broken when isolating the subsystem. As it is evident from the figure \ref{ladder}, 
the number of bonds that we break when isolating a connected subchain depends linearly on 
the size of the interval, which explains the presence of a 
linear term in the expansion of the entanglement entropy. The same is true 
for state 3, while state 1, that is the ground state of a local Hamiltonian,
does not have this property. 

\section{Fermionic chains and ladders}\label{sec7}

In this section we will try to gain physical insights from the previous results
and extend them to more general states. In order to do that, and as a continuation
of the discussion in the former section (see also \cite{Alba}), 
it will be useful
to consider $|\Psi_{\cal K}\rangle$ as the ground state of a free, homogeneous 
Hamiltonian of the following form
$$H=\sum_{n=1}^N\sum_{j=1}^{N/2} J_j a^{\dagger}_{n} a_{n+j}+ h.c.,$$
where $a_{n+N}=a_n$ and the coupling constants $J_j$ may take complex values.
Of course, due to the translational invariance of the Hamiltonian
the eigenstates are of the general form introduced in (\ref{state})
$$\vert \Psi_{\cal K}\rangle=\prod_{k\in {\cal K}} b_k^{\dagger} |0\rangle$$
with eigenvalue
$$E_{\cal K}=\sum_{k\in{\cal K}} \Lambda_k$$
where the dispersion relation is given by
$$\Lambda_k=\sum_{j=1}^{N/2}J_je^{2\pi ikj/N}+c.c.$$
The ground state of this Hamiltonian is therefore
$$|GS\rangle=\prod_{\Lambda(k) <0} b_k^{\dagger} |0\rangle$$

\begin{figure}[H]
  \centering
    \resizebox{12cm}{!}{\includegraphics{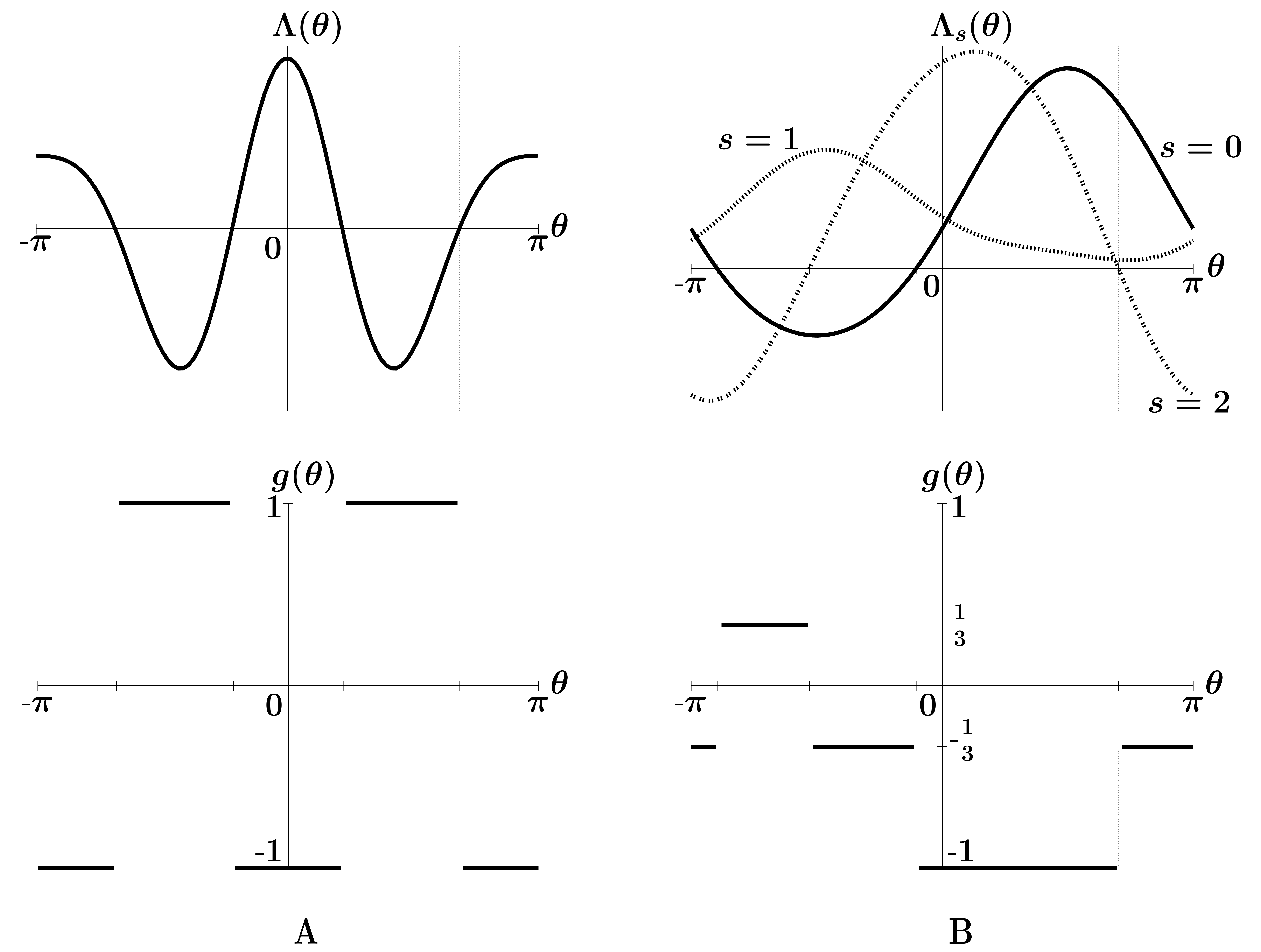}} 
    \caption{The upper left plot represents the dispersion relation 
for a local chain, 
while the occupation density for its ground state is
plotted below.
The diagrams in the right represent respectively the bands for the 
dispersion relation
of a local ladder with $3$ rails (up) and the occupation density
of its ground state (down).
}
  \label{fig-bands}
   \end{figure}

We consider two types of Hamiltonians according to the behaviour of the couplings with $j$.
We shall say that the Hamiltonian is local in the chain if the couplings are rapidly decaying in $j$, 
i.e.  $$\lim_{j\to\infty}J_j\,j^\gamma=0,\ \mbox{for any real}\ \gamma.$$ 
In this case \cite{Alba} the thermodynamic limit of the dispersion relation
$$\Lambda(\theta)=\sum_{j=1}^\infty J_j e^{i\theta j}+c.c.,$$
is a smooth function that generically has a finite (even) number of zeros 
where $\Lambda$ changes its sign, call this number $\nu$.
A typical example with $\nu=4$ is represented in 
the upper half of Fig. \ref{fig-bands} A.

The occupation density for the ground state will be given by 
\begin{equation}
g(\theta)=
\begin{cases}
1,&\mbox{if}\ \Lambda(\theta)< 0\\ 
-1,&\mbox{if}\ \Lambda(\theta)> 0,
\end{cases} 
\end{equation}
which is a piecewise constant function with values $\pm1$.
Therefore, the linear term of the entanglement entropy is absent and we get
$$S_\alpha(X) = \frac{1+\alpha}{\alpha}\frac{\nu}{12}\log L+C_\alpha,$$
where the constant $C_\alpha$ depends on the precise location 
of the discontinuities. 
We represent in Fig. \ref{fig-bands} A (lower half) 
the occupation density for the particular dispersion relation plotted
on top it.

We would like to highlight some features of this result, as they will be useful
for our later discussion. 

First we see that the logarithmic coefficient
is universal as it does not change, in general, under small changes of the couplings.
It depends solely on the changes of sign in $\Lambda$. This does not apply for 
the constant coefficient, that changes as the zeros of $\Lambda$ move.

Second, the coefficient of the logarithmic term has a simple physical interpretation:
it counts the number of massless excitations in
the thermodynamic limit of the Hamiltonian (zeros of $\Lambda$).
This result is consistent with the conformal field theory interpretation, where the central charge 
in a free field theory also counts the number of massless bosons.
On the other hand, if the Hamiltonian has a mass gap then $\Lambda$ has constant sign
and the entropy exactly vanishes. We will see how these properties can be 
translated to our second scenario: fermionic ladders.

\begin{figure}[H]
  \centering
    \resizebox{8cm}{!}{\includegraphics{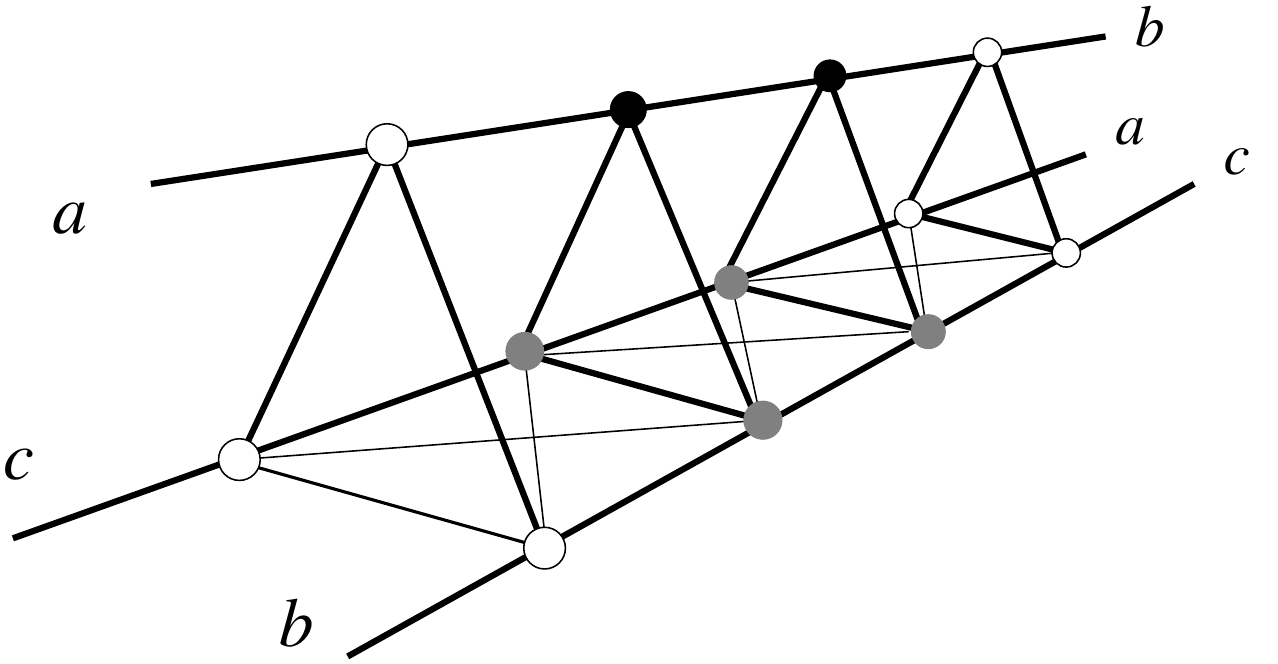}} 
    \caption{Representation of a triangular ladder corresponding to
the Hamiltonian of (\ref{ham-ladder}). The rails are joined at the end after a twist, forming therefore a twisted ring. 
For the sake of clarity,  next to nearest 
neighbours interactions are represented (with thin lines) only in the lower 
lateral face of the ladder, but they are present in all three faces.  
Subsystems are represented by dark sites: the black ones represent the 
interval while black and grey ones represent the fragment.
}
  \label{triangle}
   \end{figure}

Our second type of Hamiltonian corresponds to a prismatic ladder with $q$ rails
and local interactions. The Hamiltonian is
\begin{equation}\label{ham-ladder}
H=\sum_{n=1}^N\sum_{p=0}^{q-1}\sum_{j=0}^{N/(2q)} J_{p,j} a^{\dagger}_{n} a_{n+pN/q+j} + h.c.,
\end{equation}
where, as before, the couplings $J_{p,j}$ form rapidly decaying sequences in $j$.
An example with $q=3$ and $J_{p,j}=0$ for $j\geq 2$, is depicted in Fig. \ref{triangle}.

The dispersion relation for this system has $q$ bands, one for each value of the 
residue of $k$ (mod $q$). Namely, if $k=s$ (mod $q$) we have
$$\Lambda_k=\sum_{p=0}^{q-1} e^{2\pi isp/q}\sum_{j=0}^{N/(2q)}J_{p,j} e^{2\pi i kj/N}+ c.c.,$$
or taking the thermodynamic limit and replacing $2\pi k/N$ by the continuous 
variable $\theta$ we get
$$\Lambda_s(\theta)=
\sum_{j=0}^\infty \left(\sum_{p=0}^{q-1} e^{2\pi isp/q} J_{p,j} \right) e^{i \theta j}+c.c.$$
That is, one gets independent dispersion relations for different values of $s=0,\dots, q-1$, we call them bands.
Given the conditions for the coupling constants it is clear that every band 
is a  smooth periodic function. A generic situation with $q=3$ is 
represented in the figure \ref{fig-bands} B.
We will say that the ladder has a mass gap if the dispersion relation does 
not vanish at any point 
and we say that it is critical if some of the bands changes of sign at 
some angle 
$\theta_r$. To every change of sign we associate a massless excitation.
The example in the figure corresponds to a critical ladder with four 
massless particles.

From the dispersion relation one can deduce the ground state, where all the
modes with negative energy (Dirac sea) are occupied. Its occupation density represents the excess
of bands with negative energy over those with positive energy at a point $\theta$ 
divided by $q$. More explicitly, we define
\begin{equation}\label{gs}
g_s(\theta)=
\begin{cases}
+1,& \mbox{if}\ \Lambda_s(\theta)<0\\ 
-1,& \mbox{if}\ \Lambda_s(\theta)>0\\ 
\end{cases}
\end{equation}
and therefore the ground state for the ladder has an occupation density given by
\begin{equation}\label{occ_ladder}
g(\theta)=\frac1q\sum_{s=0}^{q-1} g_s(\theta).
\end{equation}

As an illustration, we plot in Fig. \ref{fig-bands} B the occupation density 
corresponding to 
the dispersion relation represented over it. 
In these local ladder models, the occupation density
is a piecewise constant function, like those of (\ref{occupation}), with
rational values for $t_r$.
It is also clear that every occupation density with these characteristics can be derived from 
the ground states of a local ladder Hamiltonian. 

Now we want to relate properties of the entanglement entropy of the ground state 
with those of the Hamiltonian. We will focus on two characteristics of the entropy:
the existence of linear, logarithmic and constant terms and the behaviour of the coefficients
under changes of the coupling constants.

One first property is that the linear term of the entropy is generically present.
This is not surprising because, as it was stressed in the previous section,
the correct interpretation of the {\it area law} in a ladder leads to a linear
dependence on the size of our subsystem. 

On the other hand, the logarithmic term is associated to 
discontinuities in $g(\theta)$ and it is zero if the theory has a mass gap
and, therefore, none of the bands cross the zero value.
On the contrary, the logarithmic coefficient is generically non vanishing if the theory is gapless, i.e.
when some of the bands change of sign (an exception occur in the unlikely case
in which two bands cross the zero value at the same point in opposite directions).
The constant coefficient is in general non null when the logarithmic term is present
and it is zero otherwise. 

The second aspect we look at is the universality of the coefficients.
The logarithmic term is always universal, it only depends 
on the value of the occupation density at both sides of the discontinuities 
and this 
is not affected by a small variation of the Hamiltonian. 
The linear term, however,  is not universal in general
(it depends on the position of the zeros of the
dispersion relation) but it is if the theory has a mass gap.
Then, $g(\theta)$ is constant 
and it is not affected by small changes
in the dispersion relation.
Notice that in this case the logarithmic coefficient is zero.

Finally we would like to study the physical meaning of the coefficient
of the logarithmic term. As discussed before, in the case of the local chain the interpretation 
is very simple as the term is proportional to the number of massless excitations in the theory. Here the 
situation is more involved because the coefficient of the logarithmic term depends not only on the 
number of discontinuities (massless particles) but also on the value
of the occupation density at both sides of the discontinuity.
Moreover, the relation between such values and the contribution to the logarithmic term is highly non trivial (\ref{Ba}) 
and hard to interpret.

We believe that these difficulties have the same origin that the linear term: the choice of the subsystem
$X$ as an interval in one of the rails of the ladder
(black dots in Fig. \ref{triangle}). 
Somehow,  it would have been more natural to take
for the subsystem an actual fragment of the ladder, i.e. equal intervals 
opposite to each other, one in every rail.
It is represented by black and grey sites in Fig. \ref{triangle}.  
From the point of view of the chain with non-local interactions 
this corresponds to $q$ intervals of size $L$
separated by a distance $N/q-L$. The question is if one can compute the entanglement entropy of such a
subsystem and if it has a simple physical meaning.
The answer is yes and the result is completely compatible with the interpretation of the entropy
in terms of massless particles.

Call $X_p=\{1+pN/q,\dots,L+pN/q\}$ and consider the subsystem 
$$X=\bigcup_{p=0}^{q-1} X_p$$
that actually consists of a fragment of the ladder of length $L$. The goal is to compute
the entanglement entropy of $X$ in the ground state with occupation density 
(\ref{occ_ladder}) in the thermodynamic limit.

To proceed, introduce for $s=0,\dots,q-1$
$$
(V_s)_{nm}=\frac{q}N\bigg( 
\sum_{\substack{
k\in{\cal K}\\
k=s\,(mod\,q)
}}
e^{2\pi i k(n-m)/N} 
-
\sum_{\substack{
k\not\in{\cal K}\\
k=s\,(mod\,q)
}}
e^{2\pi i k(n-m)/N}
\bigg)
. 
$$
Then the matrix
$V$ in (\ref{Vnm}) is given by
$$V=\frac1q\sum_{s=0}^{q-1} V_s.$$
On the other hand one has
$$(V_s)_{n+pN/q,m+p'N/q}=e^{2\pi i s(p-p')/q}(V_s)_{nm},$$
which implies that if we introduce the matrices
$$(T_s)_{pp'}=e^{2\pi i s (p-p')/q},\quad p,p'=0,\dots,q-1$$
we have 
$$V(X)=\frac1q\sum_{s=0}^{q-1} V_s(X_0)\otimes T_s.$$
Recall that by $V(X)$ we denote the $qL$-dimensional matrix obtained
by restricting $V$ to the indices that belong to $X$, while $X_0=\{1,\dots,L\}$
is one of the intervals that compose $X$.

Now the matrices $T_s$ commute and are diagonalised simultaneously by
$$U_{pp'}=\frac1{\sqrt{q}} e^{2\pi i pp'/q}$$ so that
$$(UT_sU^{-1})_{pp'}=q\delta_{s,p}\delta_{s,p'}.$$
As a result, taking all together we have
$$
(I\otimes U) V(X) (I\otimes U^{-1})=
\begin{pmatrix}
V_0(X_0)&0&\cdots&0\\
0&V_1(X_0)&\cdots&0\\
\cdots&\cdots&\cdots&\cdots\\
\cdots&\cdots&\cdots&\cdots\\
0&0&\cdots&V_{q-1}(X_0)
\end{pmatrix}.
$$
Where for the ground state, in the thermodynamic limit,  we have
$$(V_s)_{nm}=\frac1{2\pi}\int_{-\pi}^\pi g_s(\theta) e^{i\theta(n-m)} d\theta,$$
with $g_s$ defined in (\ref{gs}).
Therefore if we denote by $S_{\alpha,s}(X_0)$ the 
R\'enyi entanglement entropy
for the single interval $X_0$ in the state whose 
occupation density is $g_s$, for the entropy 
of the whole fragment $X$ of the ladder, we have 
$$S_\alpha(X)=\sum_{s=0}^{q-1} S_{\alpha,s}(X_0).$$
Now $S_{\alpha,s}(X_0)$ can be computed using the results of the previous section
and, if $\nu_s$ represents the number of discontinuities in $g_s(\theta)$ 
(changes of sign in $\Lambda_s(\theta)$),
we obtain
$$S_{\alpha,s}(X_0)=\frac{\alpha+1}\alpha \frac{\nu_s}{12}\log L+ C_{\alpha,s},$$
where the constant coefficient $C_{\alpha,s}$
depends on the position of the discontinuities of $g_s(\theta)$ and not only on
its number. Note that given that $g_s(\theta)=\pm1$, the linear coefficient vanishes.

Finally the logarithmic coefficient for $S_\alpha(X)$ is
$$B_\alpha=\frac{\alpha+1}\alpha \frac{\sum_s{\nu_s}}{12}.$$
Which coincides with the conformal field theory result for a number
of massless particles equal to the total changes of sign
of the $q$ bands for the dispersion relation.

\section{Conclusions and final comments}\label{sec8}

In this paper we have studied the R\'enyi entanglement entropy for a subsystem 
of a unidimensional fermionic chain in a general, translational invariant state 
that can be described by a Slater determinant. 

Although the computational  time, in principle, grows exponentially 
with the size of the subsystem we 
can circumvent this limitation and go to large sizes by employing the 
relationship between the reduced density matrix of the subsystem and the corresponding correlation
matrix. In this way, we can compute easily the R\'enyi entropy through the spectrum of the correlation matrix 
restricted to the chosen subsystem. It is important to highlight that this method is only valid
when the chain is in an state which fulfils the Wick decomposition property. Inside this set of states, there is a kind of configurations
(i.e. the factorised momentum states) in which the total correlation matrix is Toeplitz. 
In this case, if the subsystem is a single interval of contiguous sites in the coordinate space its
correlation matrix is Toeplitz as well. 

Taking the thermodynamic limit and periodic boundary conditions 
we have found the behaviour of the R\'enyi entropy with the length of the
interval, using the Fisher-Hartwig conjecture for Toeplitz determinants. 
We have computed the different coefficients
for the expansion of the entropy and we present 
explicit, general expressions for the logarithmic and constant terms,
that were not known in the literature. In order to do so we apply a new 
strategy that allows to circumvent the problem of the presence
of divergences that finally cancel out.
A simple integration by parts in the complex integral shows the automatic 
cancellation of divergences, renders the integrals finite and 
makes the computations much simpler. As remarked before, neither the
technical tool for the cancellation of divergences nor the explicit 
expression for the coefficients where previously known.

We have also checked numerically the validity of this 
expansion for three particular states. The first of them corresponds
to the ground state of a local Hamiltonian, in which case  we can compare our 
results with those derived from conformal field theory.
We have shown the perfect agreement between both approaches.
The other two states can be viewed as the lowest energy state of non-local Hamiltonians,
which is illustrated with two simple examples.
 
Finally, we discuss the physical insights that we can get from the previous 
expansions. In order to do that we study the ground state of Hamiltonians
corresponding to local chains and ladders for which the results of section 
\ref{sec5} apply.
We analyse the universality
of the coefficients and compare our results with those
derived from general arguments in conformal field theory.
In particular we show that in the case of fermionic ladders
the latter considerations can also be applied if we take a fragment, 
instead of an interval, for our subsystem.

A natural extension of our work is the study of the entropy for 
a subsystem composed of several disjoint intervals. There are some recent 
works where this problem is 
addressed for the ground state of a local chain
\cite{Caraglio, Furukawa, Calabrese2, Fagotti}. 
We would like to consider a general energy eigenstate like those in 
the present paper. 

A first step in this direction, that we took in section \ref{sec7},
is the computation of the entanglement entropy for the fragment of the ladder. 
This can be seen, actually, as a collection of disjoint intervals, but in very
particular positions so that the techniques for Toeplitz determinants and 
the Fisher-Hartwig conjecture can still be applied.

In the general case however, for intervals in generic positions, 
the reduced density matrix of the subsystem
is not Toeplitz any more, so we can not use the Fisher-Hartwig conjecture
and the present results do not apply.
The study of the expansion of the entanglement entropy for such a subsystem
will be the subject of further research.  

\noindent{\bf Acknowledgements:}  We acknowledge Germ\'an Sierra for useful
comments. Research partially supported by grants 2012-E24/2, 
DGIID-DGA and FPA2012-35453, MINECO (Spain).

\end{document}